\documentclass[twocolumn,preprintnumbers,amsmath,amssymb,prb]{revtex4}

\usepackage{graphicx}% Include figure files
\usepackage{amssymb,amsmath}
\usepackage{dcolumn}% Align table columns on decimal point
\usepackage{bm}% bold math
\usepackage{color}
\usepackage{enumerate}
\usepackage{epstopdf}

\begin{document}

\title{Floquet engineering of the Luttinger Hamiltonian}
\author{O. V. Kibis$^{1}$}\email{Oleg.Kibis(c)nstu.ru}
\author{M. V. Boev$^1$}
\author{V. M. Kovalev$^{1}$}
\author{I. A. Shelykh$^{2,3}$}

\affiliation{${^1}$Department of Applied and Theoretical Physics,
Novosibirsk State Technical University, Karl Marx Avenue 20,
Novosibirsk 630073, Russia} \affiliation{$^2$Science Institute,
University of Iceland, Dunhagi 3, IS-107, Reykjavik, Iceland}
\affiliation{$^3$Department of Physics and Engineering, ITMO University, Saint Petersburg 197101, Russia}

\begin{abstract}

Within the Floquet theory of periodically driven quantum systems,
we developed the theory of light-induced modification of
electronic states in semiconductor materials described by the
Luttinger Hamiltonian (the electronic term $\Gamma_8$).
Particularly, exact solutions of the Floquet problem are found for
the band edge in the cases of linearly and circularly polarized
irradiation. It is shown that the irradiation changes electron
effective masses near the band edge, induces anisotropy of the
electron dispersion and splits the bands. It is demonstrated that
the light-induced band splitting strongly depends on the light
polarization. Namely, the circularly polarized light acts
similarly to a stationary magnetic field and lifts the spin
degeneracy of electron branches, whereas a linearly polarized
light does not affect the spin degeneracy and only splits the
bands in the center of the Brillouin zone. The present theory can
be applied to describe electronic properties of various
semiconductor structures irradiated by an electromagnetic field in
the broad frequency range.

\end{abstract}

\maketitle

\section{Introduction}

During last years, the control of electronic parameters of
condensed-matter structures by an electromagnetic field (so-called
``Floquet engineering'' based on the Floquet theory for
periodically driven quantum systems) became the important and
established research area which resulted in the discovery of many
fundamental effects (see, e.g.,
Refs.~\onlinecite{Goldman_2014,Bukov_2015,Casas_2011,Eckardt_2015,Rahav_2003,Holthaus_2015}).
It is well known that absorption of an electromagnetic field by an
electron system takes place if only a characteristic electronic
frequency coincides with a field frequency (resonant field).
However, even if the field is non-resonant and cannot be absorbed,
it still interacts with electrons. Formally, this non-resonant
interaction is described by the nonstationary Schr\"odinger
equation with a periodically time-dependent potential. Solution of
this equation is the Floquet wave function which is periodic in
time with period of the field. Averaging all electronic
characteristics obtained with using the Floquet function over the
field period, one can construct the quantum dynamics equations for
electrons ``dressed'' by the field, which are similar to the
equations for ``bare'' electrons but depend on field parameters.
As a consequence, behavior of dressed electrons can be considered
by analogy with the behavior of bare electrons, stationary
physical parameters of which (energy spectrum, effective mass,
etc.) are renormalized by the field. Therefore, the theory of
renormalization of electronic properties of any structure by an
electromagnetic field (Floquet engineering) is based on solution
of the Floquet problem for the corresponding nonstationary
Schr\"odinger equation.

Historically, investigations of the processes of strong
interaction of electrons with an electromagnetic field, which lead
to stationary renormalization of physical properties of electronic
systems by the field, have started in the middle of 20th century.
For a long time, the main objects of the investigations were
atomic and molecular systems. In particular, the first
investigations of non-resonant interaction of electrons with a
strong field were carried out for isolated atoms and lead to
observation of the atomic energy levels shift caused by light (the
Autler-Townes effect)~\cite{Cohen_book,Autler_1955}. As to
investigations of these effects in solid state structures, they
have started with the works done by Galitsky, Goreslavsky and
Elesin~\cite{Galitskii_1970,Goreslavskii_1969}, who theoretically
predicted the existence of light-induced band gaps in energy
spectrum of semiconductors, which later were observed
experimentally~\cite{Vu_2004}. Their pioneering ideas about
light-induced modification of band electronic structure of solids
were developed later theoretically and experimentally for various
crystal
structures~\cite{Kaminski_1993,Mizumoto_2003,Srivastava_2004,Mizumoto_2006,Lignier_2007,Mizumoto_2010,Ghimire_2010}.
However, effects of electromagnetic renormalization of electronic
properties of solids were ignored as rule in the most studies for
a long time because the scattering of conduction electrons
significantly obstructs experimental investigations of them.
Situation has changed when it became possible to fabricate solid
state structures with very high charge carrier mobility and,
correspondingly, with weak electron scattering. As a consequence,
during last decade were published many works dedicated to the
Floquet engineering of various solid state structures, including
quantum
rings~\cite{Kibis_2011,Kibis_2013,Sigurdsson_2014,Kibis_2015,Hasan_2016,Kozin_2018,Kozin_2018_1},
quantum
wells~\cite{Kibis_2012,Kibis_2014,Morina_2015,Pervishko_2015,Dini_2016,Avetissian_2016,Kyriienko_2017},
topological
insulators~\cite{Wang_2013,Torres_2014,Usaj_2014,Calvo_2015,Mikami_2016,Yudin_2016,Kyriienko_2019},
graphene and related 2D
materials~\cite{Oka_2009,Kibis_2010,Syzranov_2013,Perez_2014,Kristinsson_2016,Kibis_2016,Kibis_2017,Iorsh_2017,Iurov_2019,Iurov_2020},
etc.

Among various condensed-matter structures important to both
fundamental science and device applications, it should be noted
especially those of them which are based on conventional
semiconductor materials (Si, Ge and A$_3$B$_5$ semiconductors) and
gapless semiconductors (HgTe and related materials). Particularly,
the most of modern nanostructures are fabricated with using them.
Since valence band of the conventional semiconductors and band
structure of the gapless semiconductors near the band edge (the
electronic term $\Gamma_8$ in the $\Gamma$ point of the Brillouin
zone) are described by the well-known Luttinger
Hamiltonian~\cite{Luttinger_1956}, it is necessary to develop the
consistent Floquet theory for electronic systems described by the
Hamiltonian in order to control electronic properties of the
corresponding semiconductor structures by an electromagnetic
field. The present work is dedicated to solving this theoretical
problem.

The article is organized as follows. In Section II, we solved the
Floquet problem for the nonstationary Schr\"odinger equation based
on the Luttinger Hamiltonian in the presence of an electromagnetic
field. In Section III, we calculated electron dispersion of the
Luttinger Hamiltonian modified by the field and discussed possible
experimental manifestations of the field-induced renormalization
of electronic properties. The last sections of the article contain
conclusion and acknowledgments.

\section{Model}

Let us consider a semiconductor material with the electron energy
spectrum described by the Luttinger Hamiltonian (the electronic
term $\Gamma_8$), which is irradiated by a plane electromagnetic
wave with the frequency $\omega$ and the electric field amplitude
$E_0$ (see Fig.~1). Assuming size of the irradiated semiconductor
sample along the direction of the wave propagation, $d$, to be
much larger than the interatomic distance and much less than the
wave length, $\lambda=2\pi c/\omega$, we can neglect the size
quantization of electron energy spectrum of the sample and
consider the wave field inside the sample as uniform. Then
electronic states of the irradiated semiconductor sample near the
center of the Brillouin zone (the electronic term $\Gamma_8$)
within the conventional minimal coupling approach can be described
by the time-dependent Hamiltonian,
\begin{equation}\label{Ht}
\hat{\cal{H}}(\mathbf{k},t)=\hat{\cal{H}}_L(\mathbf{k}-e\mathbf{A}(t)/\hbar),
\end{equation}
where $\hat{\cal{H}}_L(\mathbf{k})$ is the Luttinger Hamiltonian,
$\mathbf{k}=(k_x,k_y,k_z)$ is the electron wave vector, and
$\mathbf{A}(t)=(A_x,A_y,A_z)$ is the vector potential of the wave
inside the semiconductor, which periodically depends on the time,
$t$. In the present analysis, we will restrict the consideration
to the isotropic approximation of the electron dispersion in the
semiconductor. Then the Luttinger Hamiltonian takes the
form~\cite{Bir_Pikus_book,Dyakonov_1981}
\begin{equation}\label{HL}
\hat{\cal
H}_{\mathrm{L}}(\mathbf{k})=\left(\gamma_1+{5\gamma}/{2}\right)
\mathbf{k}^2-2\gamma(\mathbf{k}\mathbf{J})^2,
\end{equation}
where $\gamma=(2\gamma_2+3\gamma_3)/5$, $\gamma_{1,2,3}$ are the
Luttinger parameters, and $J_{x,y,z}$ are the $4\times4$ matrices
corresponding to the electron angular momentum $J=3/2$. To perform
calculations, it is convenient to rewrite the Hamiltonian
(\ref{HL}) as a $4\times4$ matrix in the basis of Luttinger-Kohn
wave functions, $\psi_{j_z}$, which describe four-fold degenerate
electron states of the conduction and valence band in the center
of the bulk Brillouin zone (the $\Gamma$ point), and correspond to
the four different projections of electron momentum on the $z$
axis, $j_z=\pm1/2$ and $j_z=\pm3/2$ (see for more details, e.g.,
the appendix in Ref.~\onlinecite{Kibis_2019}). In this basis, the
Hamiltonian (\ref{HL}) reads
\begin{equation}\label{LHM0}
\hat{\cal H}(\mathbf{k})=
\begin{tabular}{|c||c c c c|} \hline
${j_z}\backslash {j_z}$ & ${+3/2}$ & ${+1/2}$ & ${-1/2}$ & ${-3/2}$ \\
\hline\hline ${+3/2}$ & $F$ & $H$ & $I$ & $0$ \\
${+1/2}$ & $H^*$ & $G$ & $0$ & $I$ \\
${-1/2}$ & $I^*$ & $0$ & $G$ & $-H$ \\
${-3/2}$ & $0$ & $I^*$ & $-H^*$ & $F$ \\
\hline
\end{tabular}\,,
\end{equation}
where the matrix elements are
\begin{eqnarray}\label{LP0}
F&=&(\gamma_1+\gamma)(k_x^2+k_y^2)+(\gamma_1-2\gamma)k_z^2,\nonumber\\
G&=&(\gamma_1-\gamma)(k_x^2+k_y^2)+(\gamma_1+2\gamma)k_z^2,\nonumber\\
I&=&-\sqrt{3}\gamma(k_x-ik_y)^2,\nonumber\\
H&=&-2\sqrt{3}\gamma(k_x-ik_y)k_z.
\end{eqnarray}
%%%
In the following, we will demonstrate that electronic properties
of an irradiated semiconductor substantially depend on
polarization of the electromagnetic wave. Therefore, it is
convenient to analyze the Hamiltonian (\ref{LHM0}) for the linear
and circular polarizations separately.

{\it Linear polarization.} Let the electromagnetic wave propagates
along the $x$ axis and is linearly polarized along the $z$ axis
(see Fig.~1a). Then its vector potential inside the semiconductor
can be written as
\begin{figure}[!h]
\includegraphics[width=0.7\columnwidth]{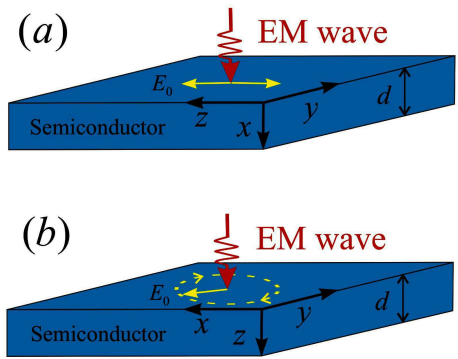}
\caption{Sketch of the semiconductor sample irradiated by an
electromagnetic wave with different polarizations: (a) linear
polarization; (b) circular polarization.}\label{Fig.1}
\end{figure}
\begin{equation}\label{AAl}
\mathbf{A}(t)=\left(0,0,\frac{E_0}{\omega}\cos\omega t\right).
\end{equation}
To simplify calculations, let us subject the Hamiltonian
(\ref{Ht}) with the vector potential (\ref{AAl}) to the unitary
transformation
\begin{equation}\label{U}
\hat{U}=\frac{1}{\sqrt{2}}
\begin{bmatrix}
e^{i\phi}&0&0&e^{i\phi}\\
0&e^{i\varphi}&e^{i\varphi}&0\\
0&-e^{-i\varphi}&e^{-i\varphi}&0\\
-e^{-i\phi}&0&0&e^{-i\phi}
\end{bmatrix},
\end{equation}
where $\phi=-3\theta/2+\pi/4$, $\varphi=-\theta/2-\pi/4$, $\theta$
is the polar angle of the electron wave vector in the $(x,y)$
plane and the electron wave vector as a function of the angle
$\theta$ reads
$$\mathbf{k}=(k_x,k_y,k_z)=\left(\sqrt{k_x^2+k_y^2}\cos\theta,\sqrt{k_x^2+k_y^2}\sin\theta,k_z\right).$$
Then the transformed Hamiltonian (\ref{Ht}), $\hat{\cal
H}^\prime(\mathbf{k},t)=\hat{U}^\dagger\hat{\cal{H}}(\mathbf{k},t)\hat{U}$,
takes the block-diagonal form,
\begin{equation}\label{bdH}
\hat{\cal H}^\prime(\mathbf{k},t)=
\begin{bmatrix}
\hat{\cal H}^{(1)}(\mathbf{k},t)&0\\
0&\hat{\cal H}^{(2)}(\mathbf{k},t)
\end{bmatrix},
\end{equation}
where the $2\times2$ matrices are
\begin{equation}\label{Hpm}
\hat{\cal H}^{(1)}(\mathbf{k},t)=
\begin{bmatrix}
\widetilde{F}&\widetilde{M}\\
\widetilde{M}^\ast &\widetilde{G}
\end{bmatrix},\,\,
\hat{\cal H}^{(2)}(\mathbf{k},t)=
\begin{bmatrix}
\widetilde{G}&-\widetilde{M}\\
-\widetilde{M}^\ast &\widetilde{F}
\end{bmatrix}
\end{equation}
and the matrix elements of the Hamiltonian are
\begin{eqnarray}\label{LP00}
\widetilde{F}&=&F+(\gamma_1-2\gamma)\left(\frac{eE_0}{\hbar\omega}\cos\omega t-2k_z\right)\frac{eE_0}{\hbar\omega}\cos\omega t,\nonumber\\
\widetilde{G}&=&G+(\gamma_1+2\gamma)\left(\frac{eE_0}{\hbar\omega}\cos\omega t-2k_z\right)\frac{eE_0}{\hbar\omega}\cos\omega t,\nonumber\\
\widetilde{M}&=&\frac{\gamma}{|\gamma|}|I|+i\frac{\gamma}{|\gamma|}\frac{|H|}{|k_z|}\left[k_z-\left(\frac{eE_0}{\hbar\omega}\right)\cos\omega
t\right].\nonumber
\end{eqnarray}

In the most general form, the nonstationary Schr\"odinger equation
for an electron in a periodically time-dependent field with the
frequency $\omega$ can be written as
$i\hbar\partial_t\psi(t)=\hat{\cal H}(t)\psi(t)$, where $\hat{\cal
H}(t+T)=\hat{\cal H}(t)$ is the periodically time-dependent
Hamiltonian and $T=2\pi/\omega$ is the field period. It follows
from the Floquet theorem that solution of the Schr\"odinger
equation is the Floquet function, $\psi(t)=e^{-i\varepsilon
t/\hbar}\varphi(t)$, where $\varphi(t+T)=\varphi(t)$ is the
periodically time-dependent function and $\varepsilon$ is the
electron (quasi)energy describing behavior of the electron in the
periodical
field~\cite{Goldman_2014,Bukov_2015,Casas_2011,Eckardt_2015,Rahav_2003,Holthaus_2015}.
The Floquet problem is aimed to find the electron energy spectrum,
$\varepsilon$. Let us solve the problem for the Hamiltonian
(\ref{bdH}).

The two Hamiltonians (\ref{Hpm}) describe the two spin-degenerated
electron states of the semiconductor with the same energy,
$\varepsilon(\mathbf{k})$. Therefore, one can consider any of the
two Hamiltonians $\hat{\cal H}^{(1,2)}(\mathbf{k},t)$ to find the
sought energy spectrum of the irradiated semiconductor,
$\varepsilon(\mathbf{k})$. For definiteness, let us restrict the
consideration by the $2\times2$ matrix Hamiltonian $\hat{\cal
H}^{(1)}(\mathbf{k},t)$. The Floquet problem with this Hamiltonian
can be solved accurately at $\mathbf{k}=0$. Namely, the
Hamiltonian $\hat{\cal H}^{(1)}(\mathbf{k},t)$ at $\mathbf{k}=0$
reads
\begin{equation}\label{H10}
\hat{\cal
H}^{(1)}_0(t)=\left(\frac{eE_0}{\hbar\omega}\right)^2\cos^2\omega
t
\begin{bmatrix}
\gamma_1-2\gamma&0\\
0 &\gamma_1+2\gamma
\end{bmatrix}.
\end{equation}
Exact solutions of the Schr\"odinger equation with the Hamiltonian
(\ref{H10}), $i\hbar\partial_t\Phi_{\pm}=\hat{\cal
H}^{(1)}_0(t)\Phi_{\pm}$, are the two Floquet functions
$\Phi_{\pm}=\exp(-i\varepsilon_\pm t/\hbar)\phi_\pm$, where
\begin{eqnarray}\label{Phi}
\phi_-&=&\begin{bmatrix}
e^{-i[({\gamma_1-2\gamma})/{4\hbar\omega}]\left({eE_0}/{\hbar\omega}\right)^2\sin2\omega t}\\
0
\end{bmatrix},\nonumber\\
\phi_+&=&\begin{bmatrix}
0\\
e^{-i[({\gamma_1+2\gamma})/{4\hbar\omega}]\left({eE_0}/{\hbar\omega}\right)^2\sin2\omega
t}
\end{bmatrix},
\end{eqnarray}
are the eigenspinors of the Floquet problem with the Hamiltonian
(\ref{H10}) and the corresponding electron energies at
$\mathbf{k}=0$ are
\begin{equation}\label{E12}
\varepsilon_\pm=\frac{\gamma_1\pm2\gamma}{2}\left(\frac{eE_0}{\hbar\omega}\right)^2.
\end{equation}

Eqs.~(\ref{Phi})--(\ref{E12}) describe the exact solutions of the
Floquet problem with the Luttinger Hamiltonian at the band edge
($\mathbf{k}=0$) in the case of linearly polarized irradiation. As
to the Floquet problem for $\mathbf{k}\neq0$, it can be solved
approximately for small electron wave vectors, $\mathbf{k}$, as
follows. In the new orthonormal basis (\ref{Phi}), the Hamiltonian
$\hat{\cal H}^{(1)}(\mathbf{k},t)$ reads
\begin{equation}\label{H1r}
\hat{\cal H}^{(1)}(\mathbf{k},t)=
\begin{bmatrix}
\overline{F}&\overline{M}\,\,\\
\overline{M}\,^\ast &\overline{G}\,\,
\end{bmatrix},
\end{equation}
where its matrix elements are
\begin{eqnarray}\label{LP1}
\overline{F}&=&\varepsilon_-+F-2(\gamma_1-2\gamma)k_z\left(\frac{eE_0}{\hbar\omega}\right)\cos\omega t,\nonumber\\
\overline{G}&=&\varepsilon_++G-2(\gamma_1+2\gamma)k_z\left(\frac{eE_0}{\hbar\omega}\right)\cos\omega t,\nonumber\\
\overline{M}&=&\widetilde{M}e^{-i\eta\sin2\omega t},\nonumber
\end{eqnarray}
and $\eta=(\gamma/{\hbar\omega})({eE_0}/{\hbar\omega})^2$. To find
the energy spectrum, $\varepsilon(\mathbf{k})$, for small electron
wave vectors $\mathbf{k}$, one can apply the conventional
perturbation theory for periodically driven quantum
systems~\cite{Goldman_2014,Bukov_2015,Casas_2011,Eckardt_2015,Rahav_2003}
to the time-dependent Hamiltonian (\ref{H1r}). As a result, we
arrive at the effective time-independent Hamiltonian,
\begin{equation}\label{H0eff}
\hat{\cal
H}_{\mathrm{eff}}(\mathbf{k})=\frac{1}{T}\int_0^{T}\hat{\cal
H}^{(1)}(\mathbf{k},t)dt,
\end{equation}
which is correct for small electron wave vectors, $\mathbf{k}$,
satisfying the condition $\mathbf{k}^2\ll\hbar\omega/|\gamma|$.
Using the well-known Jakobi-Anger expansion,
$e^{iz\sin\gamma}=\sum_{n=-\infty}^{\infty}J_n(z)e^{in\gamma}$, to
transform the exponential factor in the Hamiltonian (\ref{H1r}),
the effective Hamiltonian (\ref{H0eff}) can be rewritten in the
explicit form as
\begin{align}\label{Heff}
&\hat{\cal H}_{\mathrm{eff}}(\mathbf{k})=\nonumber\\
&\begin{bmatrix}
\varepsilon_-+F&\frac{\gamma}{|\gamma|}\left[|I|+i\frac{k_z}{|k_z|}|H|\right]J_0(\eta)\,\,\\
\frac{\gamma}{|\gamma|}\left[|I|-i\frac{k_z}{|k_z|}|H|\right]J_0(\eta)
&\varepsilon_++G\,\,
\end{bmatrix},
\end{align}
where $J_0(\eta)$ is the zeroth order Bessel function of the first
kind. Diagonalizing the effective Hamiltonian (\ref{Heff}), we
arrive at the sought electron energy spectrum near the $\Gamma$
point of the Brillouin zone,
\begin{align}\label{En}
&\varepsilon^{(\pm)}(\mathbf{k})=({\gamma_1}/{2})({eE_0}/{\hbar\omega})^2+\gamma_1k^2\pm\gamma[
(2k_z^2-k_x^2-k_y^2\nonumber\\
&+[{eE_0}/{\hbar\omega}]^2)^2+3(k_x^2+k_y^2)(k_x^2+k_y^2+4k_z^2)J_0^2(\eta)]^{{1}/{2}},
\end{align}
where the signs ``$\pm$'' correspond to the two branches of the
Luttinger Hamiltonian. It should be stressed that Eq.~(\ref{En})
correctly describes the electron energy spectrum near the band
edge ($\mathbf{k}^2\ll\hbar\omega/|\gamma|$) for any field
frequency $\omega$. Certainly, the energy spectrum (\ref{En}) in
the absence of the irradiation ($E_0=0$) exactly coincides with
the spectrum of the unperturbed Luttinger Hamiltonian (\ref{HL}),
$\varepsilon^{(\pm)}(\mathbf{k})=(\gamma_1\pm2\gamma)k^2$.

{\it Circular polarization.} Let an electromagnetic wave
irradiating a semiconductor propagates along the $z$ axis and is
circularly polarized in the $(x,y)$ plane (see Fig.~1b). Then its
vector potential inside the semiconductor can be written as
\begin{equation}\label{AAc}
\mathbf{A}=\left(\frac{E_0}{\omega}\cos\omega
t,\frac{E_0}{\omega}\sin\omega t,0\right).
\end{equation}
The Floquet problem with the Luttinger Hamiltonian (\ref{Ht}) and
the vector potential (\ref{AAc}) can be solved accurately in the
particular case of $\mathbf{k}=0$ as follows. Taking into account
the Luttinger Hamiltonian matrix (\ref{LHM0}), the considered
Hamiltonian (\ref{Ht}) with the vector potential (\ref{AAc}) at
$\mathbf{k}=0$ can be written in the block-diagonal form as
\begin{equation}\label{H000}
\hat{\cal{H}}_0=\begin{bmatrix} \hat{\cal{H}}^{(+)} & 0\\
0 & \hat{\cal{H}}^{(-)}
\end{bmatrix},
\end{equation}
where the Hamiltonian $\hat{\cal{H}}^{(\pm)}_0$ written in the
basis $\{\psi_{\pm3/2},\,\psi_{\mp1/2}\}$ reads
\begin{equation}\label{H0pm}
\hat{\cal{H}}^{(\pm)}=\left(\frac{eE_0}{\hbar\omega}\right)^2
\begin{bmatrix}
\gamma_1+\gamma
&-\sqrt{3}\gamma e^{\mp i2\omega t}\\
-\sqrt{3}\gamma e^{\pm i2\omega t} & \gamma_1-\gamma
\end{bmatrix}.
\end{equation}
Solving the nonstationary Schr\"odinger equation with the
Hamiltonian (\ref{H0pm}),
$i\hbar\partial_t{\Phi}^{(\pm)}_{1,2}=\hat{\cal{H}}^{(\pm)}{\Phi}^{(\pm)}_{1,2}$,
one can find the four exact Floquet functions,
$\Phi^{(\pm)}_{1,2}=\exp(-i\varepsilon^{(\pm)}_{1,2}
t/\hbar)\phi^{(\pm)}_{1,2}$, and the four eigenspinors of the
considered Floquet problem,
\begin{eqnarray}\label{Psi}
\phi_1^{(\pm)}&=&\begin{bmatrix}
\frac{\gamma}{|\gamma|}\sqrt{\frac{\Omega_\pm-\Delta_\pm}{2\Omega_\pm}}\,e^{\mp
i\omega t}\\
\sqrt{\frac{\Omega_\pm+\Delta_\pm}{2\Omega_\pm}}e^{\pm i\omega t}
\end{bmatrix}e^{i\omega t}, \nonumber\\
\phi_2^{(\pm)}&=&\begin{bmatrix}
\frac{\gamma}{|\gamma|}\sqrt{\frac{\Omega_\pm+\Delta_\pm}{2\Omega_\pm}}\,e^{\mp
i\omega t}\\
-\sqrt{\frac{\Omega_\pm-\Delta_\pm}{2\Omega_\pm}}e^{\pm i\omega t}
\end{bmatrix}e^{-i\omega t} ,
\end{eqnarray}
where
$\Omega_\pm=\sqrt{\Delta_\pm^2+3\gamma^2(eE_0/\hbar\omega)^4}$,
$\Delta_\pm=\gamma(eE_0/\hbar\omega)^2\mp\hbar\omega$, and the
corresponding electron energies at $\mathbf{k}=0$ are
\begin{eqnarray}\label{qEn}
\varepsilon^{(\pm)}_1&=&\gamma_1(eE_0/\hbar\omega)^2+\hbar\omega-\Omega_\pm,\nonumber\\
\varepsilon^{(\pm)}_2&=&\gamma_1(eE_0/\hbar\omega)^2-\hbar\omega+\Omega_\pm.
\end{eqnarray}

Eqs.~(\ref{Psi})--(\ref{qEn}) describe the exact solutions of the
Floquet problem with the Luttinger Hamiltonian at the band edge
($\mathbf{k}=0$) in the case of circularly polarized irradiation.
To solve the Floquet problem with the Luttinger Hamiltonian
(\ref{Ht}) and the vector potential (\ref{AAc}) at
$\mathbf{k}\neq0$, let us rewrite the Hamiltonian in the new
orthonormal basis (\ref{Psi}). For small electron wave vectors,
$\mathbf{k}$, satisfying the condition
$\mathbf{k}^2\ll\hbar\omega/|\gamma|$, one can apply the
conventional perturbation theory to the rewritten Hamiltonian in
the way discussed above for a linearly polarized field. As a
result, we arrive at the effective time-independent Hamiltonian,
$\hat{\cal H}_{\mathrm{eff}}(\mathbf{k})$, which is similar to the
Hamiltonian (\ref{H0eff}). Namely, the Hamiltonian $\hat{\cal
H}_{\mathrm{eff}}(\mathbf{k})$ is the Hamiltonian (\ref{Ht}) with
the vector potential (\ref{AAc}), which is rewritten in the basis
(\ref{Psi}) and time-averaged over the field period. In the
explicit form, the effective Hamiltonian describing the sought
electron energy spectrum, $\varepsilon(\mathbf{k})$, at small wave
vectors ($\mathbf{k}^2\ll\hbar\omega/|\gamma|$) reads
\begin{equation}\label{LHMM}
\hat{\cal H}_{\mathrm{eff}}(\mathbf{k})=\begin{tabular}{|c||c c c
c|} \hline
${\phi_j^{(\pm)}}\backslash {\phi_j^{(\pm)}}$ & ${\phi_1^{(+)}}$ & ${\phi_1^{(-)}}$ & ${\phi_2^{(+)}}$ & ${\phi_2^{(-)}}$ \\
\hline\hline ${\phi_1^{(+)}}$ & $A_+$ & $-C_+$ & $D_+$ & $0$ \\
${\phi_1^{(-)}}$ & $-C_+^{\ast}$ & $A_-$ & $0$ & $-D_-$ \\
${\phi_2^{(+)}}$ & $D_+^{\ast}$ & $0$ & $B_+$ & $C_-$ \\
${\phi_2^{(-)}}$ & $0$ & $-D_-^{\ast}$ & $C_-^{\ast}$ & $B_-$ \\
\hline
\end{tabular}\,,
\end{equation}
where the matrix elements are
\begin{eqnarray}\label{LPc}
A_\pm&=&{\frac{\Omega_\pm-\Delta_\pm}{2\Omega_\pm}}F+{\frac{\Omega_\pm+\Delta_\pm}{2\Omega_\pm}}G+\varepsilon^{(\pm)}_1,\nonumber\\
B_\pm&=&{\frac{\Omega_\pm+\Delta_\pm}{2\Omega_\pm}}F+{\frac{\Omega_\pm-\Delta_\pm}{2\Omega_\pm}}G+\varepsilon^{(\pm)}_2,\nonumber\\
C_\pm&=&\frac{\gamma}{|\gamma|}\Bigg[\sqrt{\frac{\Omega_+\mp\Delta_+}{2\Omega_+}}\sqrt{\frac{\Omega_-\pm\Delta_-}{2\Omega_-}}\nonumber\\
&-&\sqrt{\frac{\Omega_-\mp\Delta_-}{2\Omega_-}}\sqrt{\frac{\Omega_+\pm\Delta_+}{2\Omega_+}}\,\Bigg]H,\nonumber\\
D_\pm&=&\frac{\gamma}{|\gamma|}{\frac{\Omega_\pm\mp\Delta_\pm}{2\Omega_\pm}}I.
\end{eqnarray}
%%%
Correspondingly, the energy spectrum, $\varepsilon(\mathbf{k})$,
can be found as a solution of the secular equation,
\begin{equation}\label{seq}
\mathrm{det}[\hat{\cal
H}_{\mathrm{eff}}(\mathbf{k})-\mathbf{1}\varepsilon(\mathbf{k})]=0,
\end{equation}
where $\mathbf{1}$ is the unity matrix. To find the spectrum in
the broad range of electron wave vectors, the equation (\ref{seq})
should be solved numerically. However, the spectrum
$\varepsilon(\mathbf{k})$ can be written in analytical form for
the high-symmetry directions in the Brillouin zone,
$\mathbf{k}=(0,0,k_z)$ and $\mathbf{k}=(k_x,k_y,0)$. Namely, we
can write the spectrum as the four branches,
$\varepsilon^{(\pm)}_{1,2}(k_z)$ for $k_x=k_y=0$ and
$\varepsilon^{(\pm)}_{1,2}({k_x,k_y})$ for $k_z=0$, where
\begin{align}\label{bands1}
&\varepsilon^{(\pm)}_1({k_z})=\gamma_1\left(\frac{eE_0}{\hbar\omega}\right)^2+\hbar\omega-\Omega_\pm+\gamma_1k_z^2+\frac{2\gamma k_z^2\Delta_\pm}{\Omega_\pm},\nonumber\\
&\varepsilon^{(\pm)}_2(k_z)=\gamma_1\left(\frac{eE_0}{\hbar\omega}\right)^2-\hbar\omega+\Omega_\pm+\gamma_1k_z^2-\frac{2\gamma
k_z^2\Delta_\pm}{\Omega_\pm},
\end{align}
\begin{align}\label{bands2}
&\varepsilon^{(\pm)}_1({k_x,k_y})=\gamma_1\left(\frac{eE_0}{\hbar\omega}\right)^2+\gamma_1k^2+\xi_\pm\Bigg[\Bigg(\hbar\omega-\Omega_\pm\nonumber\\
&-\frac{\gamma\Delta_\pm(k_x^2+k_y^2)}{\Omega_\pm}\Bigg)^2+\frac{3\gamma^2(\Omega_\pm\mp\Delta_\pm)^2(k_x^2+k_y^2)^2}{4\Omega_\pm^2}\Bigg]^{1/2},\nonumber\\
&\varepsilon^{(\pm)}_2({k_x,k_y})=\gamma_1\left(\frac{eE_0}{\hbar\omega}\right)^2+\gamma_1k^2-\xi_\pm\Bigg[\Bigg(\hbar\omega-\Omega_\pm\nonumber\\
&-\frac{\gamma\Delta_\pm(k_x^2+k_y^2)}{\Omega_\pm}\Bigg)^2+\frac{3\gamma^2(\Omega_\pm\mp\Delta_\pm)^2(k_x^2+k_y^2)^2}{4\Omega_\pm^2}\Bigg]^{1/2},
\end{align}
and $\xi_\pm=(\hbar\omega-\Omega_\pm)/|\hbar\omega-\Omega_\pm|$.
It should be stressed that Eqs.~(\ref{bands1})--(\ref{bands2})
correctly describe the electron energy spectrum near the band edge
($\mathbf{k}^2\ll\hbar\omega/|\gamma|$) for any field frequency
$\omega$. In the absence of the irradiation ($E_0=0$), the
effective Hamiltonian (\ref{LHMM}) turns into the unperturbed
Luttinger Hamiltonian (\ref{LHM0}) and the solution of the secular
equation (\ref{seq}) exactly coincides with the unperturbed
electron dispersion,
$\varepsilon^{(\pm)}(\mathbf{k})=(\gamma_1\pm2\gamma)k^2$.

\section{Results and Discussion}

As it was mentioned above, the Luttinger Hamiltonian (\ref{HL})
can describe both valence band of conventional semiconductors (if
the two quantities, $\gamma_1\pm2\gamma$, are of the same sign)
and the band structure of gapless semiconductors near the band
edge (if they are of opposite signs) since the both cases
correspond to the same electronic term $\Gamma_8$ in the Brillouin
zone center~\cite{Bir_Pikus_book}. For definiteness, let us
restrict the consideration by the cases of valence band of such a
conventional semiconductor as GaAs
($\gamma_1=-6.96\,\hbar^2/2m_0$, $\gamma_2=-2.06\,\hbar^2/2m_0$,
$\gamma_3=-2.93\,\hbar^2/2m_0$)~\cite{Adachi_book} and the gapless
semiconductor HgTe ($\gamma_1=15.6\,\hbar^2/2m_0$,
$\gamma_2=9.6\,\hbar^2/2m_0$,
$\gamma_3=8.6\,\hbar^2/2m_0$)~\cite{Adachi_2004}, where $m_0$ is
the electron mass.
\begin{figure}[!h]
\includegraphics[width=1\columnwidth]{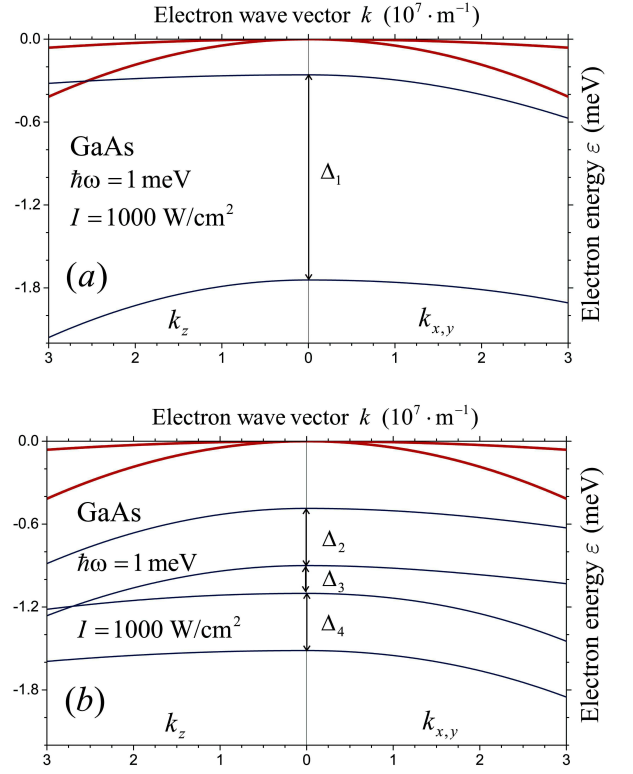}
\caption{Electron energy spectrum of GaAs sample,
$\varepsilon(\mathbf{k})$, without irradiation (red heavy lines)
and in the presence of an electromagnetic wave with the intensity
$I=1000$~W/cm$^2$, photon energy $\hbar\omega=1$~meV and different
polarizations (blue thin lines): (a) linear polarization along the
$z$ axis; (b) circular polarization in the $(x,y)$
plane.}\label{Fig.2}
\end{figure}
\begin{figure}[!h]
\includegraphics[width=1\columnwidth]{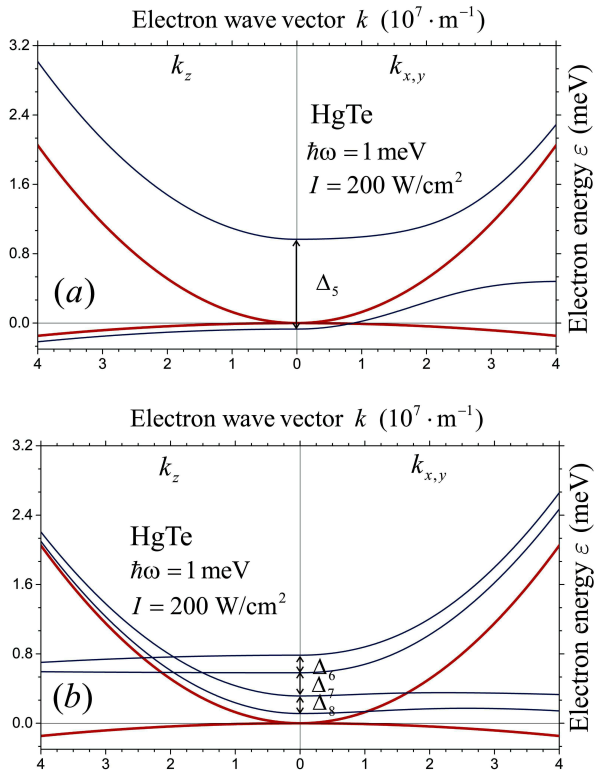}
\caption{Electron energy spectrum of HgTe sample,
$\varepsilon(\mathbf{k})$, without irradiation (red heavy lines)
and in the presence of an electromagnetic wave with the intensity
$I=200$~W/cm$^2$, photon energy $\hbar\omega=1$~meV and different
polarizations (blue thin lines): (a) linear polarization along the
$z$ axis; (b) circular polarization in the $(x,y)$
plane.}\label{Fig.3}
\end{figure}

The energy spectrum of the term $\Gamma_8$ is defined by
Eqs.~(\ref{En}) and (\ref{bands1})--(\ref{bands2}) and plotted for
GaAs (Fig.~2) and HgTe (Fig.~3) irradiated by an electromagnetic
wave with different polarizations. In the absence of irradiation,
the electronic term $\Gamma_8$ consists of the two branches which
correspond to the bands of heavy and light holes in GaAs (see the
red heavy lines in Fig.~2) and the conduction and valence bands in
HgTe (see the red heavy lines in Fig.~3). These branches are
degenerated at $\mathbf{k}=0$ and, in addition, their electron
states are doubly degenerated in spin at any electron wave vector
$\mathbf{k}$. It follows from the plots that the irradiation lifts
the degeneracy but the lifting strongly depends on the light
polarization. Namely, a linearly polarized wave splits the
electron bands at $\mathbf{k}=0$ but does not lift the spin
degeneracy of the bands (see the two blue thin lines in Figs.~1a
and 2a), whereas a circularly polarized wave lifts also the spin
degeneracy at any electron wave vector (see the four blue thin
lines in Figs.~2b and 3b). It follows from Eqs.~(\ref{En}) and
(\ref{bands1})--(\ref{bands2}) that the light-induced band
splittings marked in the Figs.~2--3 as $\Delta_{i}$ read
\begin{align}\label{Delta}
&\Delta_{1,5}=2|\gamma|\left({eE_0}/{\hbar\omega}\right)^2,\nonumber\\
&\Delta_{2,4,6,8}=\sqrt{[\gamma(eE_0/\hbar\omega)^2-\hbar\omega]^2+3\gamma^2(eE_0/\hbar\omega)^4}\nonumber\\
&+\sqrt{[\gamma(eE_0/\hbar\omega)^2+\hbar\omega]^2+3\gamma^2(eE_0/\hbar\omega)^4}-2\hbar\omega,\nonumber\\
&\Delta_{3}=2\hbar\omega-2\sqrt{[\gamma(eE_0/\hbar\omega)^2+\hbar\omega]^2+3\gamma^2(eE_0/\hbar\omega)^4},\nonumber\\
&\Delta_{7}=2\hbar\omega-2\sqrt{[\gamma(eE_0/\hbar\omega)^2-\hbar\omega]^2+3\gamma^2(eE_0/\hbar\omega)^4}.
\end{align}
Since the band splittings (\ref{Delta}) are of meV scale for the
irradiation intensities around $I\sim$~kW/cm$^2$, they can be
observed experimentally in optical electron transitions induced by
another weak (probing) electromagnetic wave. Particularly, such
optical transitions between the split bands will lead to fine
structure of the optical spectra. Besides the conventional optical
measurements, the modern angle-resolved photoemission spectroscopy
(ARPES) can also be applied to study the electron energy spectra
plotted in Figs.~2--3. Indeed, ultraviolet laser-based ARPES
provides sub-meV resolution (see, e.g.,
Refs.~\onlinecite{Zhang_book,Kiss_2005}), which is enough for
detecting features of them. It should be noted also that the band
splittings (\ref{Delta}) appear from exact solutions of the
Floquet problem at $\mathbf{k}=0$ and, therefore, go beyond the
scope of the known simple model~\cite{Rebane_1984} based on the
direct time-averaging of the Luttinger Hamiltonian.

To clarify physical nature of the light-induced band splitting, it
should be noted that a circularly polarized electromagnetic wave
breaks the time-reversal symmetry (since the time-reversal turns
left-polarized photons into right-polarized ones and vice versa).
Therefore, a circularly polarized electromagnetic wave acts
similarly to a magnetic field which lifts the spin degeneracy and
induced the asymmetry of electronic properties along the field
direction and perpendicularly to the field. As to a linearly
polarized electromagnetic wave, it acts similarly to an uniaxial
mechanical stress along the direction of polarization vector,
which both splits the degeneracy of electron states at
$\mathbf{k}=0$ and induced the anisotropy of electron
dispersion~\cite{Bir_Pikus_book}. As a consequence, the
light-induced band splitting is accompanied by the anisotropy of
electronic properties. Indeed, the unperturbed electron
dispersion,
$\varepsilon^{(\pm)}(\mathbf{k})=(\gamma_1\pm2\gamma)k^2$, is
isotropic, whereas an irradiation results in the anisotropy of
electron dispersions (\ref{En}) and (\ref{bands1})--(\ref{bands2})
along different axes in the $\mathbf{k}$ space (see the blue thin
lines in Figs.~2--3). Certainly, the anisotropy of electron
dispersion will result in the anisotropy of electron transport
which is discussed in the following.

Let charge carriers fill only ground band of the split bands near
the band edge. Then the anisotropic transport can be described by
effective electron masses. Expanding the electron energy spectra
(\ref{En}) and (\ref{bands1})--(\ref{bands2}) into the series
expansion in powers of electron wave vector $\mathbf{k}$, they can
be easily rewritten near the band edge in the parabolic form,
\begin{equation}\label{parab}
\varepsilon(\mathbf{k})=\frac{\hbar^2(k_x^2+k_y^2)}{2m_\perp}+\frac{\hbar^2k_z^2}{2m_{\|}},
\end{equation}
where $m_\perp$ and $m_{\|}$ are the electron effective masses of
the band. It should be stressed that the anisotropic electron
dispersion (\ref{parab}) takes also place in valence band of
conventional semiconductors under uniaxial mechanical stress.
Therefore, we can apply the approach known from the theory of
strained semiconductors, which is based on the the relaxation time
approximation~\cite{Bir_Pikus_book}. Within this approach, the
conductivity tensor is
\begin{equation}\label{sigma}
\sigma_{\alpha\beta}=e^2\nu\int\frac{d^3\mathbf{k}}{(2\pi)^3}v_\alpha(\mathbf{k})v_\beta(\mathbf{k})\tau(\varepsilon)\left[-\frac{\partial
f_0(\varepsilon)}{\partial\varepsilon}\right],
\end{equation}
where $\varepsilon(\mathbf{k})$ is the electron energy spectrum
(\ref{parab}) in the ground band,
$\mathbf{v}(\mathbf{k})={\partial
\varepsilon(\mathbf{k})}/{\hbar\partial\mathbf{k}}$ is the
electron velocity, $\tau(\varepsilon)$ is the electron relaxation
time, $f_0(\varepsilon)$ is the Fermi-Dirac distribution function,
and $\nu$ is the factor of spin degeneracy of the band ($\nu=1,2$
for the cases of linear and circular polarizations, respectively).
Assuming that the temperature is $T=0$, Eq.~(\ref{sigma}) yields
the sought light-induced anisotropy of conductivity,
\begin{equation}\label{anisot}
\frac{\sigma_{zz}}{\sigma_{xx}}=\frac{\sigma_{zz}}{\sigma_{yy}}=\frac{m_{\perp}}{m_\|}.
\end{equation}
Particularly, for GaAs irradiated by a linearly polarized
electromagnetic wave and the Fermi electron wave vector satisfying
the condition $k_F\ll eE_0/\hbar\omega$, the anisotropy
(\ref{anisot}) does not depend on the field parameters and reads
$$
\frac{\sigma_{zz}}{\sigma_{xx}}=\frac{\sigma_{zz}}{\sigma_{yy}}=\frac{\gamma_1-2\gamma}{\gamma_1+\gamma}.
$$

The theoretical studies of periodically driven systems described
by the Luttinger Hamiltonian, which were based on approximate
solutions of the Floquet problem, should also be
noted~\cite{Zhang_2018,Ghorashi_2018,Ghorashi_2020}. Particularly,
light-induced modifications of the electron bands originated from
the Luttinger Hamiltonian were analyzed there as a series
expansion in powers of $1/\omega$. As a consequence, these results
can be applied to real condensed-matter structures only in the
high-frequency limit. In contrast to them, we found exact
solutions of the Floquet problem with the Luttinger Hamiltonian at
the band edge. Therefore, the present theory accurately describes
all features of the electron dispersion near the band edge --- the
light-induced band splitting, the light-induced anisotropy of
effective masses, the effects depending on the light polarization,
etc
--- in the broad frequency range, including low frequencies as
well. To clarify the great importance of the low-frequency range
for condensed-matter structures, it should be noted that the
light-induced shift of electron energies in semiconductor
materials is $\Delta\sim(eE_0)^2/m^\ast\omega^2$, where  $m^\ast$
is the effective mass of charge carriers. To induce the
experimentally observable shift $\Delta\sim$~meV for infrared
frequencies ($\hbar\omega\sim$~eV), an experimental sample should
be irradiated by an electromagnetic wave with the giant intensity
$I\sim$~GW/cm$^2$, which can destruct the sample. However, the
same energy shift $\Delta\sim$~meV can be achieved with a
microwave electromagnetic field ($\hbar\omega\sim$~meV) with the
reasonable intensity $I\sim$~kW/cm$^2$. Therefore, the
low-frequency irradiation is preferable from experimental
viewpoint to observe the light-induced electronic features.
Unfortunately, the microwave frequency range goes beyond the
high-frequency approximation of the Floquet problem studied before
and needs a special consideration. As a consequence, the
low-frequency Floquet engineering of various condensed-matter
structures became the exciting research area in the
state-of-the-art Floquet theory (see, e.g.,
Ref.~\onlinecite{Vogl_2020}). Therefore, efforts to find solutions
of the Floquet problem with the Luttinger Hamiltonian in the broad
frequency range fit well the current tendencies in the
condensed-matter physics.

It should be noted also that an analysis of the electronic term
$\Gamma_8$ based on the tight-binding Hamiltonian is also
possible. However, there is the native problem of all
tight-binding Hamiltonians: The correct choice of basic (atomic)
wave functions. Since interatomic matrix elements of the
tight-binding Hamiltonian strongly depend on chosen atomic basis,
it is not an easy task to compare the theoretical calculations
with experimental data. On the contrary, the Luttinger Hamiltonian
depends only on the three parameters $\gamma_{1,2,3}$ which are
accurately found from many optical and transport measurements.
Therefore, we believe that the approach based on the Luttinger
Hamiltonian is preferable to study electronic states arisen from
the $\Gamma_8$ term.

Finalizing the discussion, it should be reminded that the present
theory is developed under assumption of continuous electron wave
vector, $\mathbf{k}$. However, the effective Hamiltonians
$\hat{\cal H}_{\mathrm{eff}}(\mathbf{k})$ derived above can also
be used to describe electronic properties of nanostructures, where
the electron wave vector is discontinuous. To take into account
the size quantization in nanostructures, one have to analyze the
Schr\"odinger problem with the Hamiltonian $\hat{\cal
H}_{\mathrm{eff}}(\hat{\mathbf{k}})+U(\mathbf{r})$, where
$\hat{\mathbf{k}}=-i\partial/\partial{\mathbf{r}}$ is the electron
wave vector operator, $U(\mathbf{r})$ is the quantizing potential
of the nanostructure, and the Hamiltonian $\hat{\cal
H}_{\mathrm{eff}}(\hat{\mathbf{k}})$ results from the effective
Hamiltonians (\ref{Heff}) and (\ref{LHMM}) with the replacement
$\mathbf{k}\rightarrow\hat{\mathbf{k}}$ .

\section{Conclusion}

Applying the Floquet formalism to electron states described by the
Luttinger Hamiltonian, we developed the theory of optical control
of the states originated from the electronic term $\Gamma_8$
(valence band in conventional semiconductors like GaAs and the
valence and conduction bands in gapless semiconductors like HgTe).
As a main result, exact solutions of the Floquet problem at the
band edge are found and the electron energy spectrum of such
materials renormalized by light is derived near the band edge for
any field frequency. It follows from analysis of the spectrum that
the electronic properties crucially depend on the irradiation
which can induce the anisotropy of electronic properties for
different directions in the Brillouin zone, band gaps in the
spectrum and the spin splitting of the bands. Possible
manifestations of the found electronic features in optical spectra
and transport measurements are discussed. Since semiconductor
materials described by the Luttinger Hamiltonian are actively used
in the modern nanotechnology, the present theory can be helpful to
describe electronic properties of various micro- and
nanostructures.

\begin{acknowledgments}
The reported study was funded by RFBR (project number
20-02-00084). I.A.S. and O.V.K. thank Icelandic Science Foundation
(project ``Hybrid polaritonics''). I.A.S. thanks the Ministry of
Science and Higher Education of the Russian Federation (Megagrant
number 14.Y26.31.0015).
\end{acknowledgments}

\end{document}